\documentstyle[prd,aps,preprint,psfig]{revtex}

\def\4he{\,{{}^4{\rm He}}}
\def\li7{\,{{}^7{\rm Li}}}

\def\mev{\,{\rm MeV}}
\def\gev{\,{\rm GeV}}
\def\MeV{\,{\rm MeV}}
\def\GeV{\,{\rm GeV}}
\def\sec{\,{\rm sec}}
\def\mb{\, {\rm mb}}
\begin{document}

\newcommand{\dw}{{\rm DW}}
\newcommand{\cw}{{\rm CW}}
\newcommand{\ml}{{\rm ML}}
\newcommand{\lt}{\tilde{\lambda}}
\newcommand{\lh}{\hat{\lambda}}
\newcommand{\phidot}{\dot{\phi}}
\newcommand{\phicl}{\phi_{cl}}
\newcommand{\adot}{\dot{a}}
\newcommand{\phat}{\hat{\phi}}
\newcommand{\ahat}{\hat{a}}
\newcommand{\hhat}{\hat{h}}
\newcommand{\phihat}{\hat{\phi}}
\newcommand{\Nhat}{\hat{N}}
\newcommand{\hth}{h_{th}}
\newcommand{\hbh}{h_{bh}}
\newcommand{\gsim}{\gtrsim}
\newcommand{\lsim}{\lesssim}
\newcommand{\bfx}{{\bf x}}
\newcommand{\bfy}{{\bf y}}
\newcommand{\bfr}{{\bf r}}
\newcommand{\bfk}{{\bf k}}
\newcommand{\bkp}{{\bf k'}}
\newcommand{\order}{{\cal O}}
\newcommand{\beq}{\begin{equation}}
\newcommand{\eeq}{\end{equation}}
\newcommand{\beqa}{\begin{eqnarray}}
\newcommand{\eeqa}{\end{eqnarray}}
\newcommand{\mpl}{M_{Pl}}
\newcommand{\lmk}{\left(}
\newcommand{\rmk}{\right)}
\newcommand{\lkk}{\left[}
\newcommand{\rkk}{\right]}
\newcommand{\lnk}{\left\{}
\newcommand{\rnk}{\right\}}
\newcommand{\call}{{\cal L}}
\newcommand{\calr}{{\cal R}}
\newcommand{\half}{\frac{1}{2}}
\newcommand{\kc}{\kappa\chi}
\newcommand{\bkc}{\beta\kappa\chi}
\newcommand{\gkc}{\gamma\kappa\chi}
\newcommand{\gbkc}{(\gamma-\beta)\kappa\chi}
\newcommand{\dchi}{\delta\chi}
\newcommand{\dphi}{\delta\phi}
\newcommand{\dOmega}{\delta\Omega}
\newcommand{\Phibd}{\Phi_{\rm BD}}
\newcommand{\echi}{\epsilon_\chi}
\newcommand{\ephi}{\epsilon_\phi}
\newcommand{\zk}{z_k}
\newcommand{\msolar}{M_\odot}
\newcommand{\mbh}{M_{\rm BH}}
\newcommand{\bh}{{\rm BH}}
\newcommand{\calf}{{\cal F}}
\newcommand{\gtilde}{{\tilde g}}
\newcommand{\Ndot}{\dot{N}}
\newcommand{\Ebar}{\overline{E}}

\tighten
\draft
\title{Primordial Black Holes and Primordial Nucleosynthesis I: \\ 
Effects of Hadron Injection from Low Mass Holes}
\author{K. Kohri}
\address{Research Center for the Early
Universe, Faculty of Science, The University of Tokyo, Tokyo 113-0033,
Japan}
\author{Jun'ichi Yokoyama} 
\address{Department of Earth and Space Science, Graduate School of
Science, Osaka University, Toyonaka, 
560-0043, Japan}

\maketitle
\begin{abstract}
    We investigate the influence of hadron injection from evaporating
    primordial black holes (PBHs) in the early stage of the primordial
    nucleosynthesis era ($t \simeq 10^{-3} - 10^4 \sec$).  The emitted
    quark-antiquark pairs or gluons immediately fragment into a lot of
    hadrons and scatter off the thermal plasma which is constituted by
    photons, electrons and nucleons. For the relatively low mass holes
    we point out that the dominant effect is the inter-conversion
    between ambient proton and neutron through the strong interaction
    induced by the emitted hadrons. Even after the freeze-out time of
    the week interactions between neutron and proton, more neutrons
    are produced and the synthesized light element abundances could be
    drastically changed.  Comparing the theoretical predictions with
    the observational data, we constrain the PBH's density and their
    lifetime. We obtain the upper bound for PBH's initial mass
    fraction, $\beta \lesssim 10^{-20}$ for $10^8 {\rm g}\lesssim M
    \lesssim 10^{10}$g, and $\beta \lesssim 10^{-22}$ for $\ 10^{10}
    \mbox{g} \lesssim M \lesssim 3\times 10^{10}$g.
\end{abstract}



\section{Introduction}
\label{sec:introduction}

Primordial black holes (PBHs) are formed in the hot early Universe if
overdensity of order of unity exists and a perturbed region enters the
Hubble radius \cite{PBH}.  They serve as a unique probe of primordial
density fluctuations on small scales.  For example, if massive compact
halo objects (MACHOs) \cite{MACHO} found by gravitational microlensing
observation toward the Large Mageranic Clouds turn out to be the PBHs,
which originate from density fluctuations generated during
inflation, we can determine model parameters of the inflation model
with high accuracy \cite{JY}.  In fact, this possibility is attracting
more attention these days because none of the conventional candidates
of MACHOs are plausible from various astrophysical considerations and
observations.

Even the nonexistence of PBHs over some mass ranges, however, would
provide useful informations on the primordial spectrum of density
fluctuations.   In this sense it is very important to obtain accurate 
constraints on the abundance of PBHs on each mass scale.

Cosmological constraints on the mass spectrum of the PBHs can be
classified to three classes.  The first one applies to heavy black
holes with mass $M > 4\times 10^{14}$g which have not evaporated by
now.  Current mass density of such holes should not exceed the total
mass density of the universe.  More stringent constraints are obtained
from the microlensing experiments on holes with sub-solar masses
\cite{MACHO}.  The second class is due to the radiation of high energy
particles from evaporating black holes \cite{evaporate,carre}.
Various constraints have been imposed from primordial big-bang
nucleosynthesis (BBN) \cite{VN,MS,ZSKC,VDN,Lindley}, microwave
background radiation \cite{cmb}, and gamma-ray background radiation
\cite{gamma}.  Finally, there may be yet another class of constraints
if PBHs do not evaporate completely but leave relics with mass of
order of the Planck mass or larger \cite{relic}.  Their mass density
should remain small enough.

The purpose of this paper is to reanalyze the effects of evaporating
primordial black holes on BBN in order to improve constraints on their
mass spectrum.  Although extensive work was done on this issue in
1970s just after the idea of PBH was proposed, to our knowledge, it
has not been studied almost two decades now, and old constraints are
still used in the literature \cite{BC,GL}.  In order to compare the
previous work with the modern view of BBN and high energy physics, we
start with a brief review of what has been done on the subject.

Vainer and Nasel'skii \cite{VN} studied the effects of the injection
of high-energy neutrinos and antineutrinos which change the epoch of
the freeze out of the weak interactions and also the neutron-to-proton
ratio at the onset of the nucleosynthesis.  This results in the
increase of $\4he$.  Demanding that its primordial abundance, $Y_p$,
should satisfy $Y_p < 0.33$, they concluded the ratio of the energy
density of PBHs to that of baryons should be smaller than $10-10^4$
for holes with mass $M=10^9-3\times 10^{11}$g.  This approximately
corresponds to $\beta(M) < 10^{-22}-10^{-19}$, where $\beta(M)$ is the
initial fraction of PBHs with mass $M$ to the total energy density of
the universe when the horizon mass is equal to $M$, {\it i.e.} 
$\beta(M) \equiv
 \left(\rho_{\bh}/ \rho_{tot}\right)_i$.  These authors also studied
the effects of entropy generation from PBHs but they obtained a modest
constraint that the energy density of the relevant holes should be
smaller than that of the photons at evaporation.

More detailed numerical analysis of the latter effect was done by
Miyama and Sato \cite{MS}, who calculated the effect of entropy
production from PBHs with mass $M=10^9-10^{13}$g evaporating during or
after the nucleosynthesis.  If these PBHs were too abundant, the
baryon-to-entropy ratio at the nucleosynthesis should be increased
which would result in overproduction of $\4he$ and underproduction of
D.  They demanded that $Y_p < 0.29 $ and that mass fraction of D
should be larger than $1\times 10^{-5}$.  If we approximate their
result with a single power law, we find \beq \beta(M) <
10^{-15}M_{10}^{-5/2}, \eeq for $M=10^9-10^{13}$g, where $M_{10}\equiv
M/10^{10}$ g.

Zel'dovich et al \cite{ZSKC}, on the other hand, studied the effect of
emission of high-energy nucleons and antinucleons from PBHs.  They claimed 
such emission would increase deuterium abundance due to capture of free
neutrons by protons and spallation of $\4he$ by emitted particles.
Their conclusions were
\beqa
 \beta (M) <&
 6\times10^{-18}M_{10}^{-1/2},~~~&{\rm for}~
 M=10^9-10^{10}{\rm g},\nonumber \\
 <&6\times 10^{-22}M_{10}^{-1/2},~~~&{\rm for}~
 M=10^{10}-5\times 10^{10}{\rm g},\nonumber \\
 <&3\times 10^{-23}M_{10}^{5/2},~~~&{\rm for}~
 M=5\times10^{10}-5\times10^{11}{\rm g},\label{constraint} \\
 <&3\times10^{-21}M_{10}^{-1/2},~~~&{\rm for}~
 M=10^{11}-10^{13}{\rm g}.\nonumber 
\eeqa

Vainer, Dryzhakova, and Nasel'skii \cite{VDN} then performed numerical
integration of nucleosynthesis network in the presence of
neutron-antineutron injection, taking the neutron lifetime to be
$918(\pm 14)$ sec and using the observational data $Y_p=0.29\pm 0.04$
and the mass fraction of D to be $5\times 10^{-5}$.
 They calculated spallation of $\4he$ and
resultant extra production of D assuming that PBHs were produced out of
scale-invariant density fluctuations.  As a result they obtained the
constraint $\beta < 10^{-26}$ for the relevant mass range.

Lindley \cite{Lindley} considered another effect of evaporating PBHs
with mass $M> 10^{10}$g, namely, photodissociation of deuterons produced
in the nucleosynthesis.  He found the constraint 
$\beta \lsim 3\times 10^{-20}M_{10}^{1/2}$, and concluded that
photodestruction was comparable to the extra production of
deuterons discussed in \cite{ZSKC}.

Since these papers were written, both observational data of light
elements and the neutron lifetime have changed considerably
\cite{BBN}.  More important, however, is the change in our view of
high energy physics of the energy scale relevant to PBHs evaporating
in the nucleosynthesis era.  That is, hadrons are not emitted in the
form of nucleons or mesons but as a quark-gluon jet in the modern view
of quantum chromodynamics (QCD).  Thus we should adopt the Elementary
Particle Picture of Carr \cite{carre} and assume that elementary
particles in the standard model are emitted from PBHs and they
generate jets.  In fact, the number of jet-generated particles far
exceeds that of directly emitted counterparts.  In this sense the
previous calculation of BBN in the presence of evaporating PBHs should
be entirely revised.  The above procedure has already been taken in
the analysis of cosmic rays from evaporating PBHs \cite{CM}.  We
incorporate emission of various hadrons from PBH-originated jets to
BBN for the first time.
 
We adopt a simple and conventional view that PBHs are produced with a
single mass, $M$, when the horizon mass is equal to $M$, namely at
$t=2.4 \times 10^{-29} M_{10} \sec \equiv t_{form}$,
  and obtain an improved constraints
on their initial abundance at each mass scale.  
Recent numerical calculations of PBH formation, however, have revealed
the mass spectrum of PBHs spreads rather widely even if the spectrum of
primordial density fluctuation is sharply peaked on a specific scale
\cite{NJ,SS}.  In particular, the authors of \cite{NJ} discovered the
critical behavior and its cosmological consequences have been discussed
in \cite{cc}.  By convolving our results with the mass functions
obtained in these papers, one can obtain constraints on PBH formation in
more realistic situations.

When PBHs evaporate and emit  various particles in BBN epoch, $i.e.$
at $T$ = 10MeV $-$ 1keV, such high energy particles interact with the
ambient photons, electrons and nucleons and they finally transfer all
the kinetic energy into the thermal bath through the electro-magnetic
interaction or the strong interaction. Through the above
process, such dangerous high energy particles possibly induce the
various effects on the background and change the standard
scenario considerably.

Once  quark-antiquark pairs
or gluons are emitted from a 
PBH, a lot of hadrons, {\it e.g.} pions,
kaons and nucleons (protons and neutrons) are produced through their
hadronic fragmentation. They inter-convert the ambient protons and
neutrons each other through the strong interaction.  If the
inter-conversion rate between neutrons and protons becomes large again
after the freeze-out time of the weak interactions in the standard BBN
(SBBN) or $t \simeq 1 \sec$, the protons which are more abundant 
than the neutrons at that time are changed into neutrons.  
That is, this results in an excess of neutrons compared with SBBN.
  Therefore, the 
hadron injection significantly influences the freeze-out value of the
neutron-to-proton ratio and the final abundances of 
$^4$He, D and $^7$Li
 are drastically changed.  In this case it is expected that
both  $^4$He and D tend to become more abundant than in 
SBBN.  If PBHs are so massive that they continue to emit particles
even after $\4he$ are produced, then spallation of $\4he$ due to high
energy particles will tend to  increase the final D
\footnote{This situation has been studied in \cite{Dim} in the context
of late-decaying particles.}.  In this
paper, as a first step of the full analysis, we concentrate on the
 low-mass PBHs and consider only the former effects.  The latter issue 
will be addressed in a separate publication~\cite{kyII}.

Reno and Seckel~\cite{RS} investigated the detail of
the physical mechanism and the influences of the hadron
injection from  long-lived massive decaying particles on BBN.  They
constrained  the parent particle's lifetime and the number density
comparing the theoretical prediction of the light element abundances
with the observational data.  Here we basically follow
 their treatment and apply
 it to the hadron injection  originated in the PBH evaporation.

The rest of the paper is organized as follows.  In \S II, we review
properties of  black hole evaporation and jets.  In \S III formulation
to calculate the effect of hadron injection to BBN is given and
observational data of light element abundances are summarized.
The result is presented in \S IV.  Finally \S V is devoted to the
conclusion.  We use the units $c=\hbar=k_B=1$.

\section{Evaporation and Jets}
\label{sec:evapolation}

First we briefly summarize basic results of PBH evaporation. As was
first shown by Hawking \cite{evaporate}, a black hole with mass $M$
emits thermal radiation with the temperature given by 
\beq
T_{\bh}=\frac{1}{8\pi GM}=1.06 M_{10}^{-1} {\rm TeV}.  \eeq 
More
precisely, a neutral non-rotating black hole emits particles with
energy between $Q$ and $Q+dQ$ at a rate 
\beq d\Ndot_s = \frac{\Gamma_s
dQ}{2\pi}\frac{1}{e^{(Q/T_{\bh})}-(-1)^{2s}}, 
\eeq 
where $s$ is the
spin of the emitted particle.  $\Gamma_s$ is its dimensionless
absorption coefficient whose functional shape is found in \cite{Page}.
It is related with the absorption cross section $\sigma_s(M,Q)$ as
$\Gamma_s(M,Q)=Q^2\sigma_s(M,Q)/\pi $.  In the high-energy limit $Q
\gg T$, $\sigma_s$ approaches to the geometric optics limit
$\sigma_g\equiv 27\pi G^2M^2$.

The average energies of a neutrino, an electron, and a photon are
given by $E_\nu=4.22T_{\bh}$, $E_e=4.18T_{\bh}$, and
$E_\gamma=5.71T_{\bh}$, respectively.  The peak energy of the flux and
that of the power are within 7\% of the above values \cite{MW}.
Averaging over degrees of freedoms of quarks and gluons
 in the standard model, we find the
average energy $\Ebar=4.4T_{\bh}$.

MacGibbon elaborated on the lifetime of a PBH, $\tau_{\bh}$, summing up the
contribution of all the emitted particles and integrating the mass loss
rate
\beq
  \frac{dM_{10}}{dt}=-5.34\times 10^{-5}f(M)M_{10}^{-2} \sec^{-1},
\eeq
over the lifetime \cite{Mac}.  The result is approximately given by
\beq
   \tau_{\bh}=6.24\times 10^3 f(M)^{-1}M_{10}^3 \sec .
\eeq
Here $f(M)$ is a function of the number of emitted particle species
normalized to unity for $M \gg 10^{17}$g holes which emit only photons,
three generations of neutrinos and anti-neutrinos, and gravitons.
Depending on the spin $s$, each relativistic degree of freedom
contributes to $f(M)$ as \cite{Mac}
\beqa
 f_{s=0}=0.267,~~~f_{s=1}=0.060,~~~f_{s=3/2}=0.020,~~~f_{s=2}=0.007, 
  \\ \nonumber
 f_{s=1/2}=0.147~ {\rm (neutral)},~~~f_{s=1/2}=0.142~
 {\rm (charge}=\pm e).
\eeqa
Summing up contributions of all the particles in the standard model, we
find $f(M)=14.34$.  In this case the lifetime reads
\begin{equation}
    \label{eq:lifetime}
    \tau_{\bh} = 435M_{10}^3  \sec ,
\end{equation}
or
\begin{equation}
    \label{eq:mass_pbh}
    M = 1.32 \times 10^{9}  \left( \frac{\tau_{\bh}}{1
    \sec }\right)^{\frac13} {\rm g}. 
\end{equation}
Thus a PBH evaporating by the end of BBN $t\simeq
10^3$ sec
 has a mass $M \lesssim 10^{10}$g and the temperature $T_{\bh} \gtrsim
1$TeV.

For such a PBH with temperature higher than the QCD scale, $\Lambda_{\rm
QCD}\sim 10^{2.5}$MeV,
it has been argued \cite{MW} that particles  radiated from it can be
regarded as asymptotically free at emission.  The emitted quarks and
gluons fragment into further quarks and gluons until they cluster into
the observable hadrons when they have traveled a distance $\Lambda_{\rm
QCD}^{-1}\sim 10^{-13}$cm.  The hadron jet thus produced would be
similar to that produced in $e^+ e^-$ annihilation.

We now estimate the average number of the emitted hadron
species $H_i$ per jet as
\begin{equation}
    \label{eq:NHi}
    N^{H_i} =  f_{H_i} \frac{\langle N_{ch}\rangle}{2} ,
\end{equation}
where $\langle N_{ch}\rangle$ is the averaged
charged-particle multiplicity  which represents the
total number of the  charged particles emitted per two hadron jets,
$f_{H_i}$ is the number fraction of the hadron species $H_i$ to all the
emitted charged particles. 

\begin{figure}
  \begin{center}
    \centerline{\psfig{figure=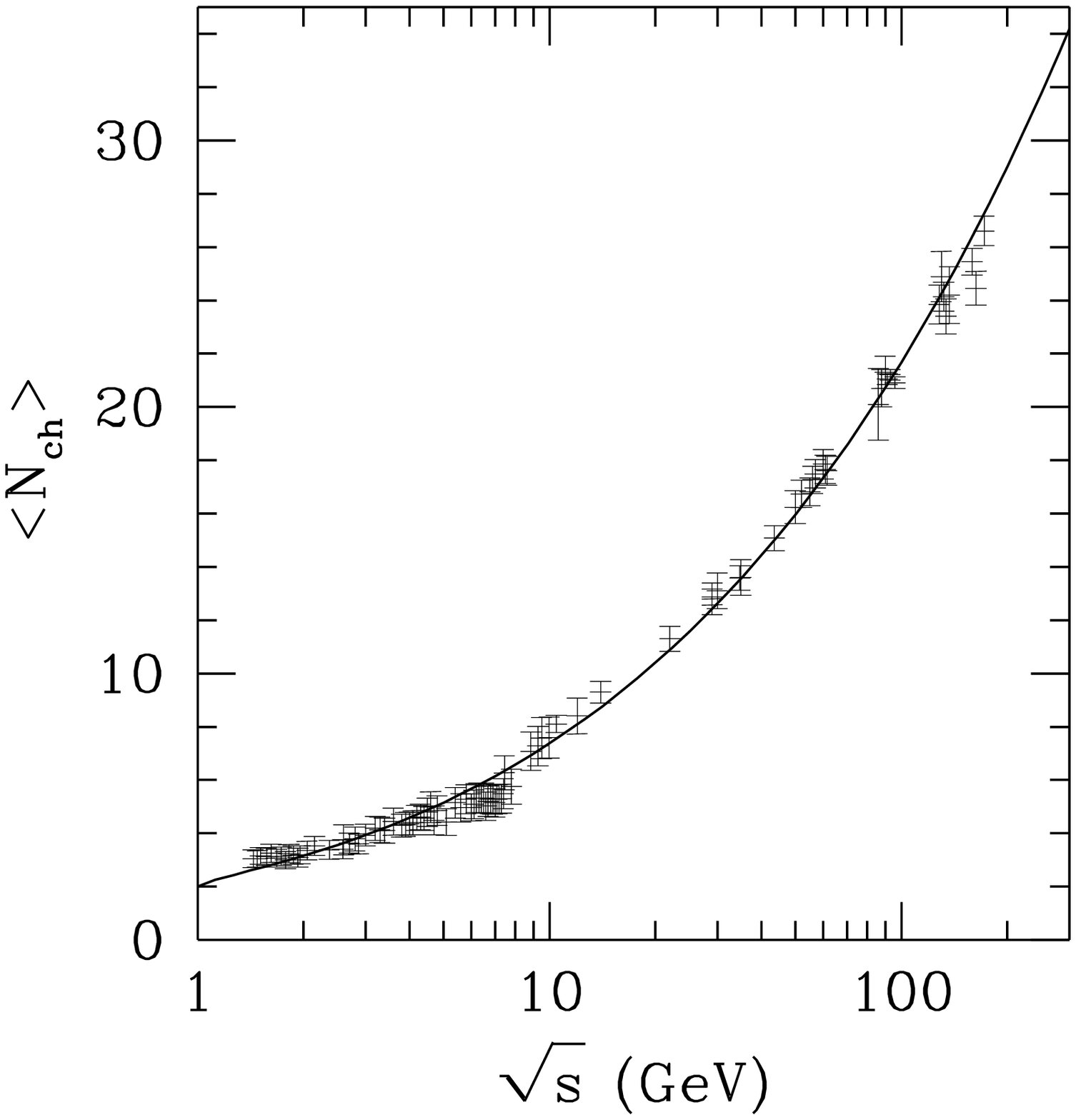,width=16cm}}
      \caption{
      Plot of the charged particle multiplicity $\langle
      N_{ch}\rangle$ for the center of mass energy $\sqrt{s}$. At
      about $ \sqrt{s} \simeq 91.2$ GeV the experimental data are
      spread horizontally by authors. We assigned systematic errors of
      10$\%$ for $\sqrt{s} = 1.4 - 7.8$ GeV .}
      \label{fig:nch}
  \end{center}
\end{figure}

It is  reasonable to assume that the averaged charged-particle
multiplicity is independent of the source of the hadron jet. In this
paper we use the data obtained by
the $e^+e^-$ collider experiments and extrapolate it to higher
energy scales. A number of experimental data are available at least up
to $\sqrt{s} \simeq$
100 GeV where $\sqrt{s}$ denotes the center of mass energy.  Recently
LEPII experiments (ALEPH, DELPHI, L3 and OPAL) yielded  useful data
for $\sqrt{s}$ = 130 $-$ 172 GeV.  
One can fit the data by a function,
 $\langle N_{ch}\rangle=a + b
\exp(c\sqrt{\ln(s/\Lambda^2)})$, where $\Lambda$ is the cut-off
parameter in the perturbative calculations associated with the onset
of the hadronization and $a,~ b$ and $c$ are constants~\cite{QCDfit}.
The above functional shape  is
motivated by the next-to-leading order perturbative QCD calculations.
 We
fit the data of the $e^+e^-$ collider experiments for $\sqrt{s}$ = 1.4
$-$ 172 GeV~\cite{PDG} and we find
\begin{equation}
    \label{eq:nch}
    \langle N_{ch}\rangle= 1.73 + 0.268 \exp\lmk 
   1.42\sqrt{\ln(s / \Lambda^2)}\rmk,
\end{equation}
with $\chi^2$ = 90.3 for 93 data points and the error of the fitting
is about 10$\%$. Here we have taken $\Lambda = 1$ GeV. We assigned
systematic errors of 10$\%$ for $\sqrt{s} = 1.4 - 7.8$ GeV because
MARK I and $\gamma \gamma 2$ experiments do not include the systematic
errors in their data, see~\cite{PDG}. In Fig.~\ref{fig:nch} we plot
$\langle N_{ch}\rangle$ for the center of mass energy $\sqrt{s}$. 
When
we apply the results of the $e^+e^-$ collider experiments to the case
of PBH evaporation, we take $\sqrt{s} = 2 \Ebar$ as the
energy of two hadron jets from the PBH evaporation because we assume
that two hadron jets are induced per one quark-antiquark pair emission
here. Then we can approximately estimate the average energy of an
emitted hadron species $H_i$  as $E_{H_i} \simeq
\Ebar/\langle N_{ch}\rangle$. 

The next issue is to find $f_{H_i}$.  In this paper we adopt the
experimental data of $f_{H_i}$ at $\sqrt{s}$ = 91.2 GeV, which is the
highest energy for which these data are available \cite{PDG}, and we
assume that $f_{H_i}$ do not change significantly in the energy range
$\sqrt{s} \simeq $ 100 GeV - 20 TeV. \footnote{The above energy range
is approximately the PBH's temperature  corresponding to the
lifetime $\tau_{\bh}$ = $10^{-1}$ - $10^4$ sec. For $\sqrt{s} \gtrsim$
350 GeV, $t$-$\bar{t}$ pairs would be produced and they would change
the form of the charged particle multiplicity and the hadron fraction.
Since we do not have the experimental data for such high energy
regions, we extrapolate $\langle N_{ch}\rangle$ to the higher energy
regions and we take $f_{H_i}$ as a constant.} From the table of
$f_{H_i}$ in \cite{PDG} we must estimate effective $f_{H_i}$ for each
hadron species on the relevant time scale.  That is, we must also take
into account the decay products of those particles whose lifetime is
too short to affect BBN.

Let us estimate the time scale at which we should estimate $f_{H_i}$.
The emitted hadrons do not scatter off the ambient nucleons directly.
At first the emitted high energy hadrons scatter off the background
photons and electrons because they are much more abundant than the
nucleons.  For the most part of the epoch at the cosmic time $t
\lesssim 10^4$ sec, as we can see later, it is expected that the
emitted particles are quickly thermalized and they have the kinetic
equilibrium distributions before they interact with the ambient
nucleons. In addition as we show in Sec.~\ref{sec:result}, it is
reasonable to treat the emitted hadrons to be homogeneously
distributed. Then we use the thermally averaged cross sections
$\langle\sigma v\rangle^{H_i}_{N \rightarrow N'}$ for the strong
interaction between hadron $H_i$ and the ambient nucleon $N$, where
$N$ denotes proton $p$ or neutron $n$.  For a hadron interaction
process $N + H_i \rightarrow N' + \cdot \cdot \cdot$, the strong
interaction rate is estimated by
\begin{eqnarray}
    \label{eq:gamma^i_nn}
    \Gamma^{H_i}_{N\rightarrow N'} &=& n_N \langle\sigma v\rangle^{H_i}_{N
    \rightarrow N'}  \nonumber \\ 
    &\simeq& 10^{8} \sec^{-1} f_N
    \left(\frac{\eta_i}{10^{-9}}\right)
    \left(\frac{\langle\sigma v\rangle^{H_i}_{N \rightarrow N'}}{10
    \mb} \right)
    \left(\frac{T_{\nu}}{2 \mev}\right)^3,
\end{eqnarray}
where $n_N$ is the number density of the nucleon species $N$, $\eta_i$
is the initial baryon to photon ratio ($=n_B/n_{\gamma}$ at
$T\gtrsim10 \mev$), $n_B = n_p + n_n$ denotes the baryon number
density, $f_N \equiv n_N/n_B$, and
$T_{\nu}$ is the neutrino temperature \footnote{The $T_{\nu}$
dependence comes from that the baryon number density decrease as $n_B
\propto R(t)^{-3}$ and the scale factor exactly decrease as $R(t)
\propto T_{\nu}^{-1}$ while the universe adiabatically expands. If 
large entropy is produced, the above relation must be
changed. In the hadron injection scenario, however, the large entropy
production region is severely excluded.  Therefore 
Eq.~(\ref{eq:gamma^i_nn}) is a good  estimate of the time scale
of the hadron interactions.  
}.

Thus all we need to consider are particles with lifetime larger than
$\order (10^{-8})$ sec.  The corresponding mesons are $\pi^+,~ \pi^-,~
K^+,~K^-$, and $ K_L$ and the baryons are $p$, $\overline{p}$, $n$,
and $\overline{n}$.  We have therefore calculated the final yield of
these particles per two jets, 
$n^{H_i}$, out of Table 38.1 of \cite{PDG}.
  The results are: \beqa
n^{\pi^+}=14.1,~~  n^{\pi^-}=14.1,~~\nonumber\\
n^{K^+}=1.67,~~  n^{K^-}=1.67,~~n^{K_L}=1.19 \\
n^{p}=n^{\overline{p}}=0.772,~~ n^{n}=n^{\overline{n}}=0.772.
~~\nonumber \eeqa In estimating $n^{n}$ we have assumed that the
number of neutrons emitted directly from a jet is the same as that of
protons.  Then we obtain the number fraction $f_{H_i}$ by
\begin{equation}
    \label{eq:fHi}
    f_{H_i} = \frac{n^{H_i}}{\langle N_{ch}(\sqrt{s}=91.2 \gev)\rangle}.
\end{equation}

\section{Hadron injection and BBN}
\label{sec:hadron}
\subsection{Hadron scattering off the background particles}

When there are sufficient electrons and positrons in the
universe ($T \gtrsim$ 0.025 MeV), it is expected that such charged
mesons are quickly thermalized through the Coulomb scattering.  Since
the stopping time to lose the relativistic energy is estimated as
\beq
 \tau_{ch} \simeq 10^{-14} \sec (E/\GeV)/(T/\MeV)^2
\eeq
 where $E$ is the
kinetic energy of a charged meson, the long-lived charged mesons are
thermalized and scatter off the ambient nucleons by the threshold
cross section before they decay \cite{RS}.  
The thermally averaged cross sections
for $\pi^{\pm}$ are obtained by~\cite{RS}
\begin{eqnarray}
    \label{eq:sigma_pi}
    {}&& \langle\sigma v\rangle^{\pi^+}_{n \rightarrow p} = 1.7 \ \mb, \\ 
    {}&& \langle\sigma v\rangle^{\pi^-}_{p \rightarrow n} =
    1.5C^{\pi}(T) \ \mb,
\end{eqnarray}
where $C^{H_i}(T)$ is the Coulomb correction factor. Because the
reaction $p^+ + \pi^- \rightarrow n + \cdot \cdot \cdot$ is enhanced
due to the opposite-sign charge of the initial state particles, we
correct the strong interaction rates by multiplying $C^{H_i}(T)$ to
that which are obtained by ignoring the Coulomb corrections. The
Coulomb correction factor is estimated by
\begin{equation}
    \label{eq:coulomb}
    C^{H_i}(T) = \frac{2\pi \xi_i(T)}{1 - e^{\ 2\pi \xi_i(T)}},
\end{equation}
where $\xi_i(T) = \alpha \sqrt{\mu_i/ 2T}$, $\alpha$ is the fine
structure constant and $\mu_i$ is the reduced mass of the hadron
$H_i$ and the nucleon. 

The cross sections for $K^-$ are obtained by~\cite{RS}
\begin{eqnarray}
    \label{eq:sigma_ka-}
    {}&& \langle\sigma v\rangle^{K^-}_{n \rightarrow p} = 26 \ \mb, \\ 
    {}&& \langle\sigma v\rangle^{K^-}_{n \rightarrow n} = 34 \ \mb, \\ 
    {}&& \langle\sigma v\rangle^{K^-}_{p \rightarrow n} = 31  C^{K}(T)
      \ \mb, \\ 
    {}&& \langle\sigma v\rangle^{K^-}_{p \rightarrow p} = 14.5
    C^{K}(T) \ \mb.
\end{eqnarray}
Following Reno and Seckel \cite{RS} we ignore $K^+$ interaction
because $n + K^+  \rightarrow p + K^0$ is the endothermic reaction
which has $Q = 2.8$ MeV.

On the other hand, among neutral kaons, $K_L$ has a long
lifetime $\order ({10^{-8}})$ sec. 
Since $K_L$ does not stop through the
electro-magnetic interactions, we adopt the following strong
interaction cross sections which are obtained by the initial energy
distribution in the hadron fragmentation~\cite{RS},
\begin{eqnarray}
    \label{eq:sigma_kaL}
    {}&& \langle\sigma v\rangle^{K_L}_{n \rightarrow p} = 7 \ \mb, \\ 
    {}&& \langle\sigma v\rangle^{K_L}_{n \rightarrow n} = 10 \ \mb, \\ 
    {}&& \langle\sigma v\rangle^{K_L}_{p \rightarrow n} = 7 \ \mb, \\ 
    {}&& \langle\sigma v\rangle^{K_L}_{p \rightarrow p} = 10 \ \mb.
\end{eqnarray}

As for the emitted high energy nucleons, we should treat the
thermalization process more carefully because the stopping process is
not so simple. Proton and antiproton are stable and we should 
worry about the efficiency to lose the kinetic energy at the lower
temperature. At least $T \gtrsim 0.02$ MeV, protons are quickly
thermalized through Coulomb scattering off the electrons and Inverse
Compton scattering off the photons. For $t \gtrsim 3 \times 10^3$ sec,
on the other hand, the stopping process of protons proceeds through
the nuclear collisions with the ambient protons and light 
elements. In this case such high energy protons may induce the
$^4$He fissions because $^4$He has already been synthesized at around
$t \simeq 300\sec$.  For neutron and antineutron the efficiency of the
thermalization is more severe.  Neutrons are efficiently stopped by
the electron scattering until $T \simeq 0.09$MeV. Thus the late time
emission of the high energy neutron would induce the light element
fissions.  However, in this paper we treat the neutrons as if they are
approximately thermalized in the entire parameter range because we are
primarily concerned with the effects of low-mass PBHs here.  We follow
the Reno and Seckel's treatment that a nucleon-antinucleon pair is
regarded as a meson $N\bar{N}$. Then the $N\bar{N}$ meson induces the
inter-converting reactions like $N + N\bar{N} \rightarrow N' + \cdot
\cdot \cdot$ and the thermally averaged cross sections are given
by~\cite{RS}
\begin{eqnarray}
    \label{eq:sigma_nuc}
    {}&& \langle\sigma v\rangle^{n\bar{n}}_{n \rightarrow n} =  37 \ \mb, \\ 
    {}&&  \langle\sigma v\rangle^{n\bar{n}}_{p \rightarrow n} =  28 \ \mb, \\ 
    {}&& \langle\sigma v\rangle^{p\bar{p}}_{n \rightarrow p} =  28 \ \mb, \\ 
    {}&&  \langle\sigma v\rangle^{p\bar{p}}_{p \rightarrow p} =  37 \ \mb. 
\end{eqnarray}
The above treatment may underestimate the deuterium abundance 
because it
will be produced by the hadro-dissociation of $^4$He
if PBHs are so massive that they continue to emit high-energy hadrons
even after $\4he$ are formed.
The hadron-induced 
dissociation process of the light elements will be discussed in
the next paper~\cite{kyII}. For the other mesons and baryons whose
lifetimes are much shorter, {\it e.g.}  $\pi^0 \rightarrow 2\gamma$ with
$\tau_{\pi^0} \simeq \order(10^{-16}) \sec$, they quickly decay into the
standard particles and do not influence the standard scenario.

On the other hand, for the even 
longer lifetime $\tau_{\bh} \gtrsim 10^{4}
\sec$, there is
another interesting effects on BBN. The emitted photons or charged
particles induce the electro-magnetic cascade showers and produce many
soft photons. 
Their spectrum has a cutoff at $E_\gamma^{\rm max}\simeq m_e^2/(22T)$,
where $m_e$ is the electron mass \cite{Holtmann}.
If $E_\gamma^{\rm max}$ exceeds the binding
energies of the light elements, these photons dissociate
 light elements and change their abundances.   In fact, the
energy of the photon spectrum which are produced by the PBH
evaporation at $t \gtrsim 10^{4} (10^{6}) \sec$ exceeds the deuterium
($^4$He) binding energy $B_2 \simeq 2.2$ ($B_4 \simeq 20$) MeV. In
this case  PBH's number density and the lifetime are severely
constrained by the observational data of the light element
abundances~\cite{Lindley}.

\subsection{Formulation}
As we mentioned in the previous section, the hadron emission mechanism
by the PBH evaporation at $t\lesssim 10^4$ sec induces  extra
interactions between  emitted hadrons and  ambient nucleons.
That is, they enhance the inter-converting interaction rates between
neutron and proton even after the weak interactions has already
frozen out in the standard scenario and the freeze out values of $n/p$
ratio can be increased. The time evolution of the PBH's mass is given
by
\begin{equation}
    \label{eq:mass_time}
    M(t) = \left\{
 \begin{array}{ll}
  M \left(\frac{\tau_{\bh} - t}{\tau_{\bh}}
    \right)^{\frac13} \quad & ({\rm for} \ \ t \lesssim \tau_{\bh}), \\ 
  0 \qquad \qquad \qquad \qquad & ({\rm for} \ \
    \tau_{\bh} \lesssim t ), 
 \end{array}
 \right. 
\end{equation}
where  $M$ is the initial mass of PBH
when it was formed.

Then the time evolution equations for the number density of a nucleon
$N (=p, n)$ is represented by
\begin{equation}
    \label{eq:difeqN}
    \frac{dn_N}{dt} + 3 H(t) n_N = \left[\frac{dn_N}{dt}\right]_{SBBN} 
    - B_h  J(t) \left( K_{N \rightarrow N'} - K_{N' \rightarrow N} \right),
\end{equation}
where  $H(t)$ is the cosmic
expansion rate, $[dn_N/dt]_{SBBN}$ denotes the
contribution from the standard weak interaction rates \cite{BBN} and nuclear
interaction rates, $B_h$ is the hadronic branching ratio, $J(t)$
denotes the emission rate of the hadron jet per unit time and $K_{N
\rightarrow N'}$ denotes the average number of the transition $N
\rightarrow N'$ per one hadron jet emission.  The emission rate of the 
hadron jet is estimated by
\begin{equation}
    \label{eq:jetrate}
    J(t) = \frac{n_{\bh}(t)}{\Ebar(t)}  \frac{\mbox{d}M(t)}{\mbox{d}t},
\end{equation}
where $n_{\bh}(t)$ is the number density of the PBHs.

Though PBHs  generally emit not only quarks and gluons but
also all lighter particle species than the temperature of PBH, the
emitted neutrinos, photons and the other electro-magnetic particles do
not influence the light element abundances significantly for the
relatively short lifetime. As we noted in the previous section, for
$\tau_{\bh} \lesssim 10^4 \sec$ the injection of photon or the other
electro-magnetic particles do not induce the photo-dissociation of the
light elements. On the other hand, the emitted neutrinos scatter off
the background neutrinos and produce the electron-positron pairs.
Although they also induce the electro-magnetic cascade, we should not
be worried about photo-dissociation for the same reason.  Hence
we concentrate on the effects of  hadron injection in BBN epoch.
Then the resultant upper bound on the abundance of PBHs, $\beta$,
turns out to be proportional to $B_h^{-1}$.  Below we analyze the
case $B_h=1$, because its magnitude is not precisely known, although
we expect $B_h=\order(0.1-1)$.  Anyway the constraint for other cases
with $B_h \neq 1$ can easily be obtained from the above scaling law.

The average number of the transition $N \rightarrow N'$
per jet is expressed by
\begin{equation}
    \label{eq:Knn}
    K_{N \rightarrow N'} = \sum_{H_i} N^{H_i}R^{H_i}_{N \rightarrow N'},
\end{equation}
where $H_i$ runs the hadron species which are relevant to the nucleon
inter-converting reactions, $N^{H_i}$ denotes the average number of
the emitted hadron species $H_i$ per jet  which is given by
Eq.~(\ref{eq:NHi}) and $R^{H_i}_{N \rightarrow N'}$ denotes the
probability that a hadron species $H_i$ induces the nucleon transition
$N \rightarrow N'$. The transition probability is estimated by
\begin{equation}
    \label{eq:trans_prob}
    R^{H_i}_{N \rightarrow N'} = 
     \frac{\Gamma^{H_i}_{N \rightarrow N'}}{\Gamma^{H_i}_{dec} +
     \Gamma^{H_i}_{abs}}, 
\end{equation}
where 
$\Gamma^{H_i}_{dec} = \tau_{H_i}^{-1}$ is the decay rate of the hadron
$H_i$, 
$\Gamma^{H_i}_{abs}$ is the total absorption rate of
$H_i$.  For $K_L$, which is not stopped, the decay rate is
approximately estimated by $\Gamma^{K_L}_{dec}=m_{K_L}/E_{K_L}
\tau_{K_L}^{-1}$ where $E_{K_L}$ is the averaged energy of the emitted
$K_L$.

\subsection{Observational constraints}
As we mentioned in the previous subsection, the hadron injection
increases the freeze-out value of $n/p$ ratio. The remaining
neutrons are included in the deuteriums rapidly and it burns into
$^4$He. Since the effects of the late time hadron emission tend to
increase D and $^4$He  we can constrain mass and abundance of PBHs 
 comparing the yield
with the observational light element abundances. In this subsection we
briefly review the observational data of the light elements D, $^4$He
and $^7$Li.

The primordial deuterium is measured in the high redshift QSO
absorption systems. Because it is expected that such a Lyman limit
system is not contaminated by any galactic or stellar chemical
evolution, we regard it  as a primordial component. At present we have
two class of the observational D values, Low D and High D. However,
since it is premature to determine which component is primordial, we
consider both of  two cases in this paper. Burles and Tytler
observed clouds at $z$ = 3.572 towards Q1937-1009 and at $z$ = 2.504
towards Q1009+2956 and they obtained~\cite{BurTyt}
\begin{equation}
    \label{lowd}
     \mbox{D/H} = (3.39 \pm 0.25) \times 10^{-5}:\ {\rm Low~ D},
\end{equation}
where D/H denotes the deuterium number fraction and the error is the 
1$\sigma$ value.  On the other hand, Webb et al. reported the higher
abundance in relatively low redshift absorption systems at $z$ = 0.701
towards Q1718+4807, $\mbox{D/H} \simeq \order (10^{-4})$~\cite{webb}.
Tytler et al.  also observed the same clouds and obtain the high
deuterium abundance independently~\cite{tyt_high},
\begin{equation}
    \label{highd}
     \mbox{D/H} = (0.8 - 5.7) \times 10^{-4}:\  {\rm High~ D}, 
\end{equation}
where the uncertainty is 2$\sigma$.

The primordial component of the $^4$He mass fraction $\mbox{Y}_p$ is
measured in the low metalicity extragalactic HII region.  In addition
to the primordial component, $^4$He is also produced in stars together
with Oxygen and Nitrogen through the stellar evolution mechanism.
Therefore in order to obtain the primordial component from the
observational data we should regress to the zero metalicity for the
observed $^4$He values.  Olive, Skillman and Steigman~\cite{OliSkiSte}
adopted 62 blue compact galaxies (BCG) observations and they
obtained the relatively ``Low'' abundance, $\mbox{Y}_p \simeq 0.234$.
However, Thuan and Izotov~\cite{Izo} pointed out that the effect of
the HeI stellar absorption, which was not considered properly
in~\cite{OliSkiSte}, is very important and they reported the
relatively ``High'' value, $\mbox{Y}_p = 0.245 \pm 0.004$.  Recently
Fields and Olive~\cite{FieOLi} reanalyzed the observational data and
they reported
\begin{equation}
    \label{FieOLi}
     \mbox{Y}_p=0.238 \pm (0.002)_{stat} \pm (0.005)_{syst},
\end{equation}
where the errors are the 1$\sigma$ values.

The primordial value of $^7$Li/H is measured in the Pop II old halo
stars. Since the low mass stars , $i.e.$ the low temperature stars, have
the deep convective zone, the primordial component should be
considerably depleted in the warm interiors. For the large mass stars
which have the relatively high effective surface temperature $T_{eff} \gtrsim
5500$K, it is known that the primordial components are not changed and
they have a ``plateau''of the $^7$Li/H as a function of $T_{eff}$.  On
the other hand it is also reported that $^7$Li/H abundances decrease
along with Iron fraction [Fe/H]. For the plateau stars, however, the
$^7$Li abundances are not changed with decreasing [Fe/H] for
[Fe/H]$\lesssim -1.5$.  Bonifacio and Molaro observed
41 plateau stars and reported~\cite{BonMol}
\begin{equation}
    \label{li7}
    \log_{10}(^7\mbox{Li/H}) =-9.76 \pm (0.012)_{stat} \pm (0.05)_{syst}
    \pm (0.3)_{add},
\end{equation}
where $^7$Li/H denotes $^7$Li number fraction, the errors are 1
$\sigma$ values and we take an additional systematic error $\Delta
\log_{10}(^7\mbox{Li/H})_{add} = 0.3$, for fear that we may underestimate the
stellar depletion and the production by the cosmic ray spallation.

In order to get the conservative bound for the hadron injection
 induced by PBH evaporation, we adopt the following mild
observational bounds here,
\begin{eqnarray}
    \label{eq:obs4}
    {}&& \mbox{Y}_p \le 0.252 \ (2\sigma), \\
    \label{eq:obslow2}
    {}&& \mbox{D/H}  \le 4.0 \times 10^{-5}\ (2\sigma) \quad 
      {\rm for \ Low D} \\
    \label{eq:obshigh2}
    {}&&  \mbox{D/H} \le 5.7 \times 10^{-4}\ (2\sigma) \quad  {\rm for 
    \  High D}, \\
    \label{eq:obs7}
    {}&&  3.3 \times  10^{-11} \le {}^7\mbox{Li/H} \le 9.2 \times
    10^{-10}\ (2 \sigma),
\end{eqnarray}
where we summed all the errors in quadrature.

\section{Results}
\label{sec:result}
In this section we compare the theoretical predictions of the 
light-element abundances in the hadron injection scenario with the
observational constraints. Now we have three free parameters, the
baryon to photon ratio $\eta$, the PBH's lifetime $\tau_{\bh}$, and
the initial number density of the PBH normalized by the entropy
density, $s$, $Y_{\bh}\equiv n_{\bh}/s$.  We
start the BBN calculation at the cosmic temperature $T$ = 100 MeV.
Since $\eta$ is the value at present time, the initial value $\eta_i$
should be set to an appropriate value which turns out to the present
$\eta$ after the possible entropy production due to PBH evaporation and
the photon heating due to $e^+e^-$ annihilation. As we noted in the
previous sections, the lifetime $\tau_{\bh}$ characterizes not only
the decay epoch, but also the PBH's initial mass $M$ and the initial PBH
temperature $T_{\bh}$. Here we can relate $Y_{\bh}$ to the initial
mass fraction $\beta$ by the following equation:
\begin{equation}
    \label{eq:betai-ypbh}
    \beta = 5.4 \times 10^{21} \left( \frac{\tau_{\bh}}{1
    \sec}\right)^{\frac12} Y_{\bh}.
\end{equation}

Since the hadron injection tends to increase the produced D
and $^4$He abundances in the present situation, 
the parameter range of $\eta$ is necessarily
restricted to a narrow region if  we adopt a set of the
observational upper bounds for D and $^4$He. First we discuss the case
of ``LowD'' and  take Eqs.~(\ref{eq:obs4}), ~(\ref{eq:obslow2}) and
~(\ref{eq:obs7}) as a first set of the observational constraints. In
this case the baryon-to-photon ratio is restricted in $\eta = (4.7 -
8.6) \times 10^{-10}$. 

\begin{figure}
  \begin{center}
    \centerline{\psfig{figure=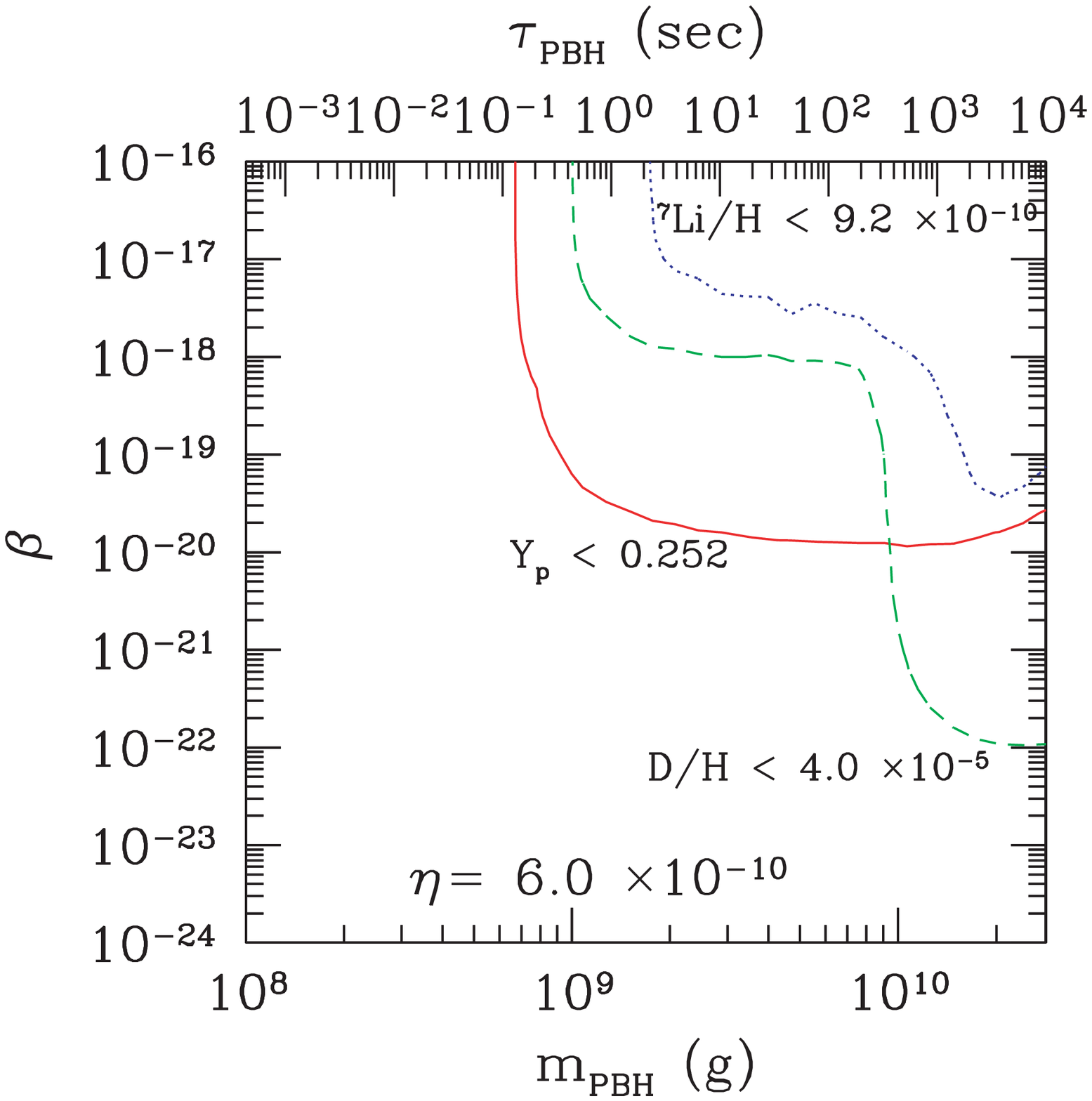,width=16cm}}
      \caption{
      Upper bounds for $\beta$ which come from the observational
      constraints of $^4$He (solid line), D (dashed line) and $^7$Li
      (dotted line) as a function of the PBH's
      mass at $\eta = 6.0 \times 10^{-10}$. Here we take ``LowD'' as a
      deuterium observational constraints.  }
      \label{fig:eta6}
  \end{center}
\end{figure}

In Fig.~\ref{fig:eta6} we plot the upper bounds for $\beta$ which
come from the observational constraints of $^4$He, D and $^7$Li for
the PBH mass $M = 10^8 - 3\times 10^{10}$g at $\eta = 6.0
\times 10^{-10}$.  The above mass range corresponds to the lifetime
$\tau_{\bh} = 4.4 \times 10^{-4} - 10^4 \sec$. As we noted earlier,
the shorter lifetime, $\tau_{\bh} \lesssim 10^{-2} \sec$, do not
affect the freeze-out value of $n/p$ and do not change any predictions
of SBBN.  However, if the lifetime becomes  $\tau_{\bh} \gtrsim
10^{-2} \sec$, the freeze-out value of $n/p$ ratio is increased by the
hadron-induced inter-converting interactions and the produced neutron
 increases the $^4$He abundance because most of the free
neutrons burned into $^4$He through D. Then PBH abundance is strongly
constrained by the upper bound of the observational $^4$He abundance.
For $\tau_{\bh} \gtrsim 10^{2} \sec$, since the produced free D can no
longer burn into $^4$He, the extra free neutrons remain in D. Then
$\beta$ is severely constrained by the upper bound of the
observational D/H. Though $^7$Li abundance traces D abundance for the
longer lifetime $\tau_{\bh} \gtrsim 10^{2} \sec$ and is produced more
than SBBN at the relatively high $\eta$ ($\gtrsim 3 \times 10^{-10}$),
the constraint is much weaker than D.

\begin{figure}
  \begin{center}
    \centerline{\psfig{figure=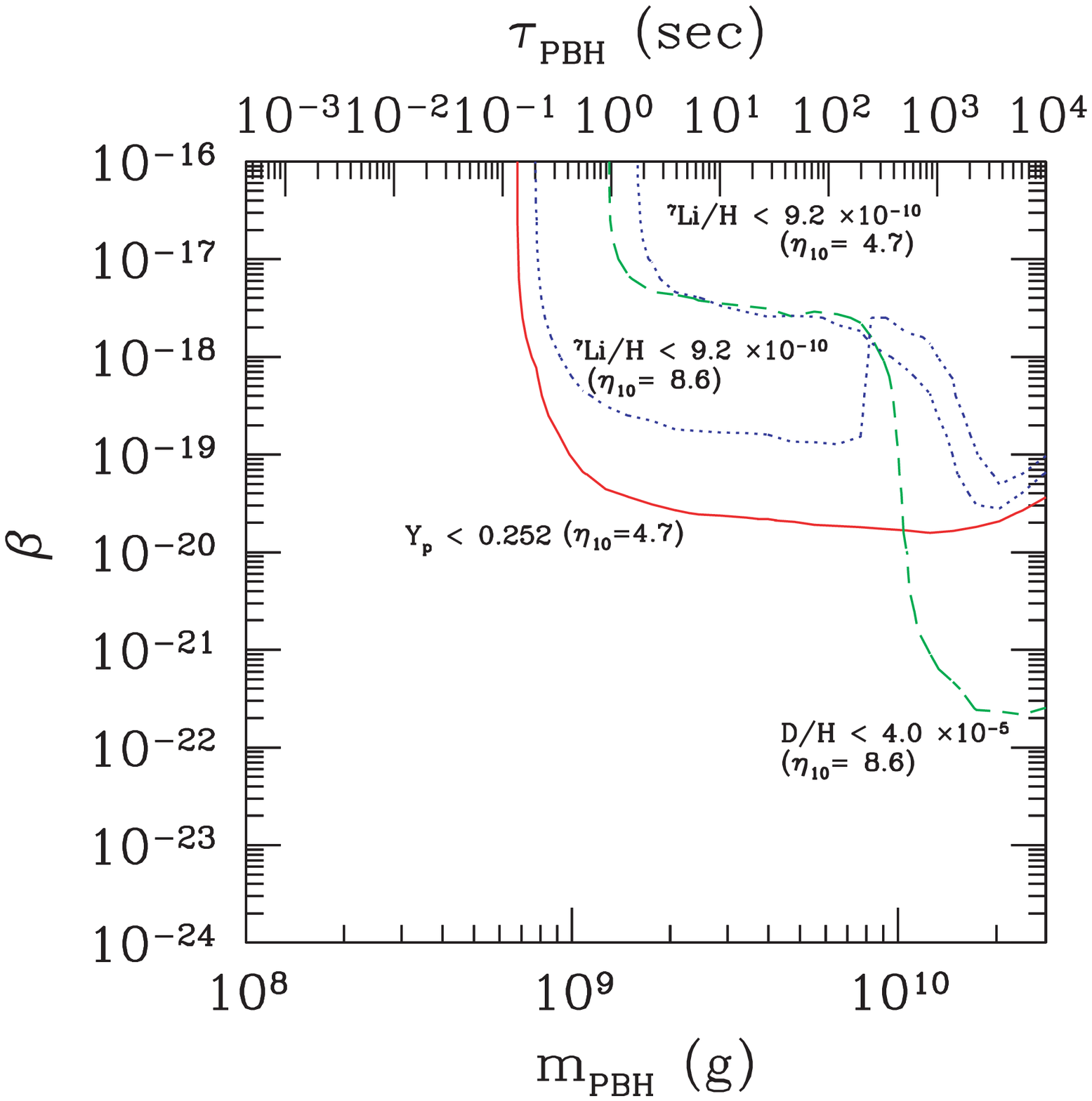,width=16cm}}
      \caption{
      The most conservative upper bounds for $^4$He (solid line), D
      (dashed line) and $^7$Li (dotted line) as mild as possible in
      the parameter region $\eta_{10} = 4.7 - 8.6$, where $\eta_{10}
      \equiv \eta \times 10^{10}$. Here we take ``LowD'' as a
      deuterium observational constraint.  }
      \label{fig:lowD}
  \end{center}
\end{figure}

In Fig.~\ref{fig:lowD} we plot the most conservative bounds 
 in the parameter region $\eta = (4.7 - 8.6)\times 10^{-10}$.
For $\tau_{\bh} \lesssim 4 \times 10^{2} \sec$ such a 
bound is obtained by $\mbox{Y}_p$ at $\eta = 4.7 \times 10^{-10}$. For
$4 \times 10^{2} \lesssim \tau_{\bh} \lesssim 10^{4} \sec$, the
prediction of D/H at $\eta = 8.6 \times 10^{-10}$ gives the most
conservative bound for $\beta$, but constraints in this region is
not quantitatively
accurate because we did not include spallation of $\4he$ produced.

\begin{figure}
  \begin{center}
    \centerline{\psfig{figure=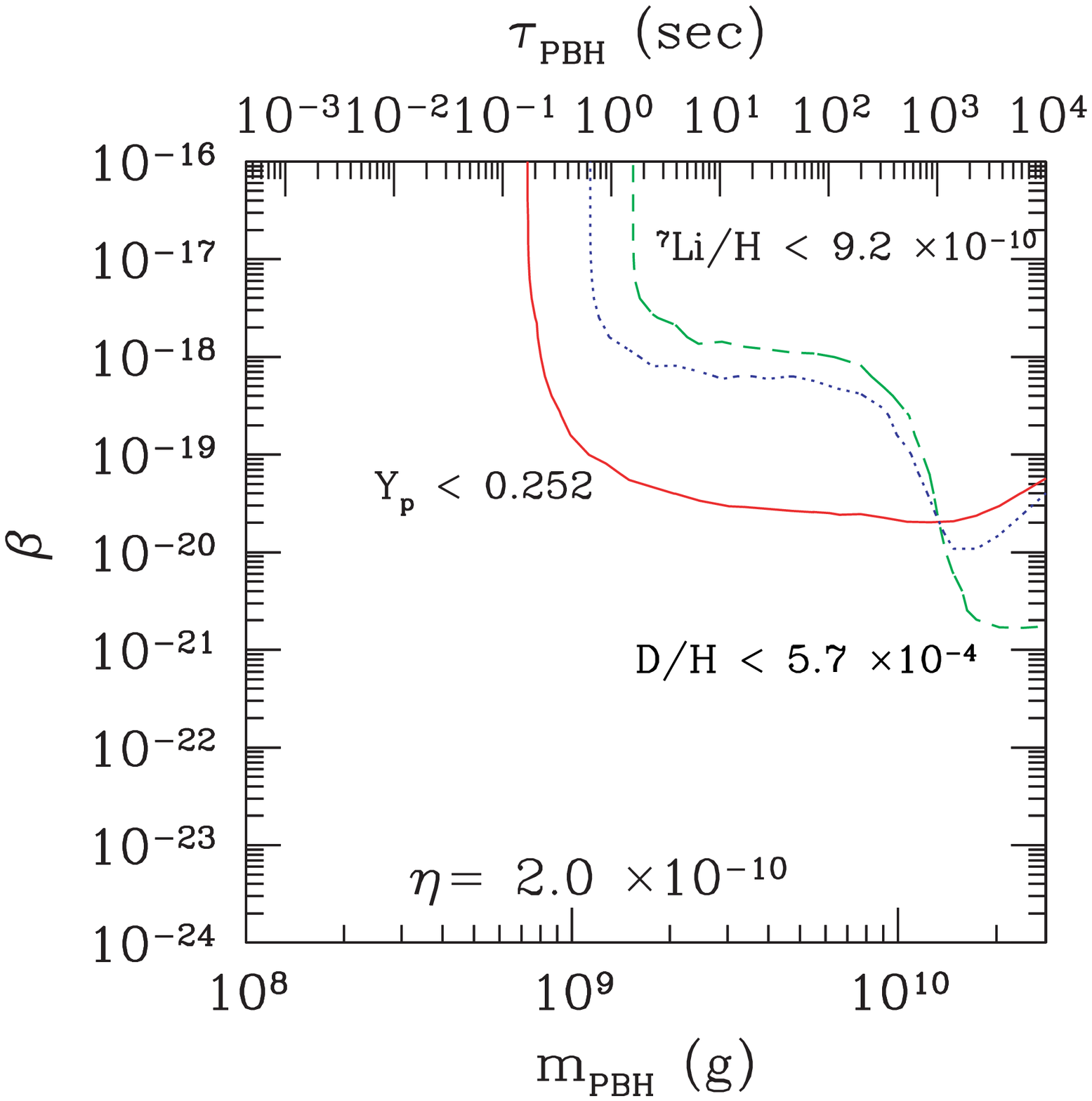,width=16cm}}
      \caption{
      Upper bounds for $\beta$ which come from the observational
      constraints of $^4$He (solid line), D (dashed line) and $^7$Li
      (dotted line) as a function of the PBH's
      mass at $\eta = 2.0 \times 10^{-10}$. Here we take ``HighD'' as
      a deuterium observational constraints. }
      \label{fig:eta2}
  \end{center}
\end{figure}

\begin{figure}
  \begin{center}
    \centerline{\psfig{figure=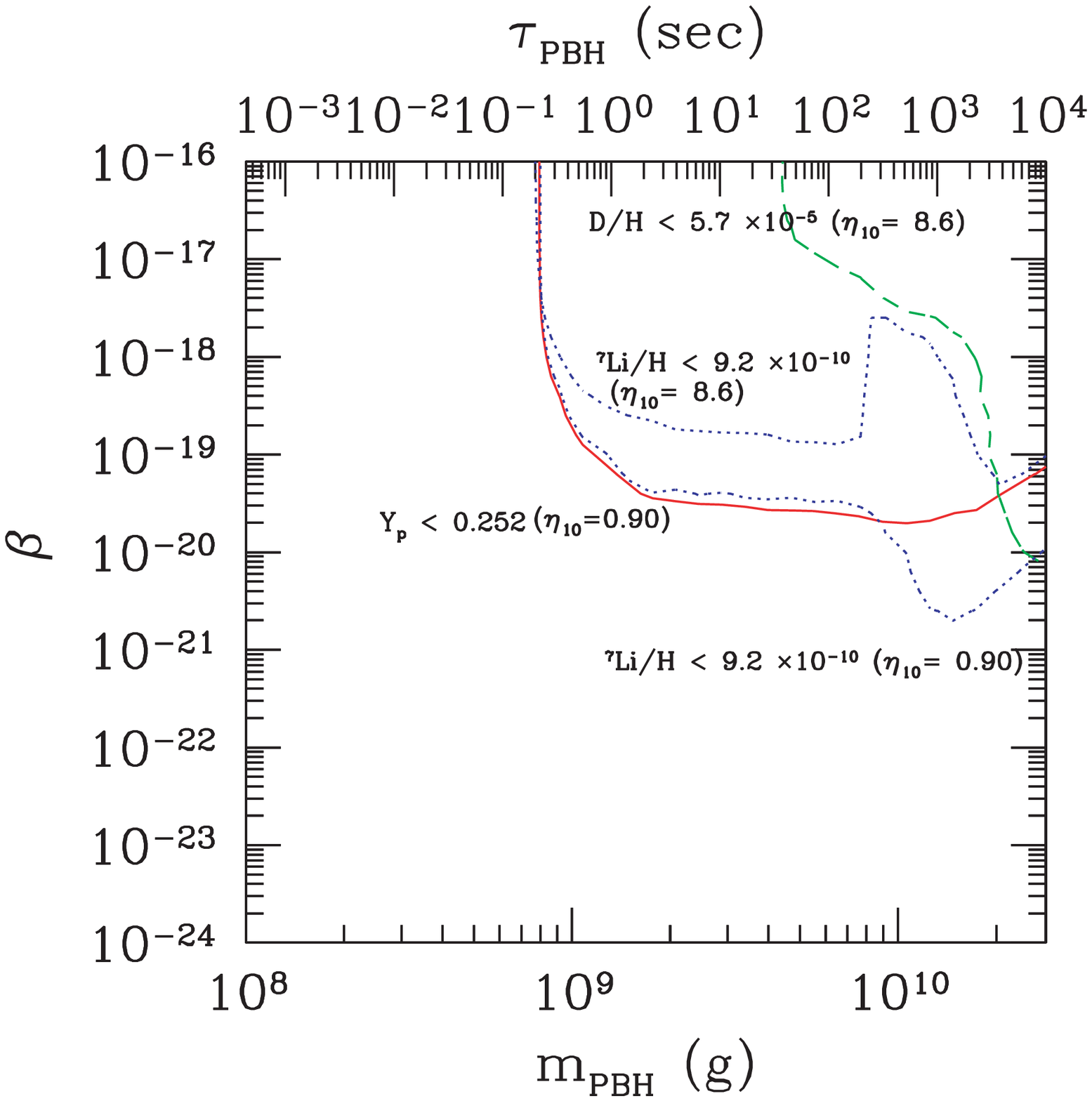,width=16cm}}
      \caption{
      The most conservative bounds for $^4$He (solid line), D (dashed
      line) and $^7$Li (dotted line) as mild as possible in the
      parameter region $\eta_{10}= 0.90 - 8.6$, where $\eta_{10}
      \equiv \eta \times 10^{10}$. Here we take ``HighD'' as a
      deuterium observational constraints.  .}
      \label{fig:highD}
  \end{center}
\end{figure}

Second we discuss the other case ``HighD'' and adopt a second set of
the observational constraints Eqs.~(\ref{eq:obs4}), 
(\ref{eq:obshigh2}), and ~(\ref{eq:obs7}). The upper bounds for D and
$^4$He restrict the baryon to photon ratio necessarily within $\eta =
(0.90 - 8.6) \times 10^{-10}$. In Fig.\ref{fig:eta2}, we plot the
upper bounds for $\beta$ as a function of PBH mass at $\eta = 2.0
\times 10^{-10}$. Because the observational upper bound for D is very
weak, the allowed region for $\tau_{\bh} \gtrsim 10^{2} \sec$ are
enlarged. Since the predicted $^4$He abundance has the log dependence
for $\eta$, the constraint which comes from $\mbox{Y}_p$ is not
changed very much compared with the case of $\eta = 6.0 \times
10^{-10}$. In Fig.\ \ref{fig:highD} we plot the most conservative upper
bounds for $\beta$ in the parameter space $\eta = (0.90 - 8.6)
\times 10^{-10}$. As we mentioned above, the D constraint is not
strict because of the weak observational upper bound. On the other
hand, the constraint from $^7$Li/H is the most severe for $\tau_{\bh}
\gtrsim 10^{2} \sec$ because $^7$Li/H are almost not allowed to be
produced at $\eta = 0.90 \times 10^{-10}$.

Finally we estimate the number of PBHs within the neutron diffusion length
$d_n(t)$ at $t = \tau_{\bh}$. The neutron diffusion length is given
by~\cite{apple_hogan}
\begin{equation}
    \label{eq:d_n}
    d_n(t) = 5.8 \times 10^2 \left( \frac{T}{1
    \mev}\right)^{-\frac52}\left( \frac{\sigma_t}{ 3 \times 10^{-30}
    {\rm cm}}\right)^{- \frac12},
\end{equation}
where $\sigma_t$ is a transport cross section. Using
Eq.~(\ref{eq:lifetime}), the number of PBHs
within the horizon length at $t =\tau_{\bh}$ is given by
\begin{eqnarray}
    \label{eq:nhorizon}
    N_{H}(\tau_{\bh})= \left(
        \frac{\tau_{\bh}}{t_{form}}\right)^{\frac32} \beta 
                      = 7.7 \times 10^{46} 
        M_{10}^3 \beta.
\end{eqnarray}

Then the number of PBHs within the neutron diffusion length is
estimated by
\begin{equation}
  N_d(\tau_{\bh}) = \left\{
 \begin{array}{ll}
 1.2 \times 10^{21} \beta &,  \quad ({\rm for} \ \tau_{\bh} = 1 \sec), \\ 
 2.3 \times 10^{46} \beta &,  \quad ({\rm for} \ \tau_{\bh} = 10^3 \sec).
 \end{array}
 \right.
 \label{eq:n_d}
\end{equation}
Compared with Figs. \ref{fig:eta6} -- \ref{fig:highD}, we can see that
there exist sufficiently many PBHs within the neutron diffusion
length in the relevant parameter spaces in which we are interested and
the number of PBHs is observationally constrained. Therefore it is
justified that we can take the emitted hadrons as a homogeneously
distributed background. 

\section{Conclusion}
\label{sec:conclusion}

We have investigated the influence of hadron injection from
 evaporating PBHs in the early stage of the BBN
 epoch ($t \simeq 10^{-3} - 10^4 \sec$). The above lifetimes correspond to the
mass range of PBHs $M \simeq 10^8 - 3
\times 10^{10} \mbox{g}$. When quark-antiquark pairs or gluons are
emitted from PBH, they immediately fragment into a lot of hadrons
(pions, kaons and little nucleons) and the produced hadron jets are
injected into the thermal plasma which is constituted by photons,
electrons and nucleons. After the emitted hadrons are sufficiently
stopped through the electro-magnetic interaction with the ambient
photons and electrons, they can scatter off the ambient nucleons and
they inter-convert proton and neutron each other. As a result
 more neutrons are produced and the synthesized light
element abundances are drastically changed. In particular, $^4$He and
D abundances are very sensitive to the neutron abundance and tend to
increase in this scenario. 
Comparing with the observational data, we can
constrain  PBH's  density and their lifetime. The hadron
injection  which are originated by the direct quark-antiquark
pairs or gluons emission from PBHs has never considered at all in the
literature. In this paper we pointed out that the hadron induced
inter-converting interactions between protons and neutrons are very
important and the energy density of PBH is severely constrained by the
observational data.

We obtain the following upper bounds for the initial mass fraction of
PBH $\beta$ as a function of the initial mass of PBH when they are
formed
\begin{eqnarray}
    \label{eq:constraints}
    \beta  &\lesssim&  10^{-20} \quad ({\rm for} \  10^8 {\rm g}
    \lesssim M \lesssim
    10^{10} \mbox{g}) \\
    \beta  &\lesssim& 10^{-22} \quad ({\rm for} \ 10^{10} \mbox{g} 
      \lesssim M \lesssim 3\times 10^{10} \mbox{g}).
\end{eqnarray}
Here we adopt the case of ``lowD'' because the produced D is more
severely constrained, and we take the most conservative bounds which
are independent of the baryon to photon ratio.

\section*{Acknowledgments}
\label{sec:acknowledgements}
We wish to thank O. Biebel for providing the experimental data of the
charged particle multiplicity and his detailed explanations of them.
We are also grateful to M.\ Kawasaki, M.M.\ Nojiri, S.\ Orito, and
N.\ Sasao for useful communications.  This work was partially
supported by the Monbusho Grant-in-Aid for Scientific Research Nos.\ 
10-04502 (KK) and 11740146 (JY).




\end{document}